\documentclass[aps,reprint,nofootinbib]{revtex4-1}

\usepackage[utf8]{inputenc}

\usepackage{graphicx}
\usepackage{dcolumn}
\usepackage{color}
\usepackage{amsmath}
\usepackage{bm}
\usepackage{hyperref}
\usepackage[export]{adjustbox}

\newcommand{\qc}{\,\text{,}}

\definecolor{forestgreen}{rgb}{0.1,0.49,0.07}

\begin{document}

\title{TDI-$\infty$: time-delay interferometry without delays}

\author{Michele Vallisneri}
\author{Jean-Baptiste Bayle}
\affiliation{Jet Propulsion Laboratory, California Institute of Technology, Pasadena CA 91109 USA}
\author{Stanislav Babak}
\author{Antoine Petiteau}
\affiliation{AstroParticule et Cosmologie (APC), Université de Paris/CNRS, 75013 Paris, France}


\begin{abstract}
The space-based gravitational-wave observatory LISA relies on a form of synthetic interferometry (time-delay interferometry, or TDI) where the otherwise overwhelming laser phase noise is canceled by linear combinations of appropriately delayed phase measurements. These observables grow in length and complexity as the realistic features of the LISA orbits are taken into account.
In this paper we outline an \emph{implicit} formulation of TDI where we write the LISA likelihood directly in terms of the basic phase measurements, and we marginalize over the laser phase noises in the limit of infinite laser-noise variance. Equivalently, we rely on TDI observables that are \emph{defined numerically} (rather than algebraically) from a discrete-filter representation of the laser propagation delays.
Our method generalizes to any time dependence of the armlengths; it simplifies the modeling of gravitational-wave signals; and it allows a straightforward treatment of data gaps and missing measurements.  
\end{abstract}


\maketitle

\section{Introduction}

Interferometry is not indispensable to the experiments that seek to detect gravitational waves (GWs) by monitoring the displacement of freely falling test masses.
Sensitivity is set by disturbances to free fall (acceleration noise) and by the precision of the distance measurement (position noise).
Interferometry becomes crucial when the ruler by which distance is measured (typically, the wavelength of an infrared laser) is not sufficiently stable at the GW frequencies of interest, so its fluctuations must be canceled out interferometrically.
Such is the case of the space GW observatory LISA \cite{lisa2017}, in which laser frequency noise is several orders of magnitude larger than acceleration and position noise. LISA is a strange kind of interferometer, where the laser-noise-canceling interferometric observables are not realized physically, but reconstituted in post-processing from the set of one- or two-way phase measurements between the pairs of spacecraft in the constellation.

This reconstitution is known as \emph{time-delay interferometry} (TDI, \cite{Tinto:1999yr,Armstrong_1999,Tinto:2002de,Tinto:2004,TintoLRR2014}) because the phase measurements are delayed by multiples of the LISA armlengths before they are combined.
While this design is ingenious, and indeed seminal to the LISA concept, it is inconvenient for data analysis. GWs are completely buried in the laser-noise-dominated phase measurements, so both the phase data and the theoretical GW templates must undergo a time-domain transformation, which is computationally costly and time dependent (because the LISA armlengths are changing continuously).
TDI compounds the difficulties of data reduction: for instance, gaps in the phase measurements are replicated multiple times across the TDI time series \cite{2019PhRvD.100b2003B}; clock noise requires a complicated subtraction procedure \cite{2020arXiv200502430H};
stretching LISA armlengths couple noisily to the interpolation of the delays \cite{Bayle:2019}; and more.

In this paper we propose that the LISA phase measurements can be analyzed \emph{directly} for the purpose of GW detection and parameter estimation, without transforming them explicitly into TDI observables with analytical forms derived \emph{a priori}.
Equivalently, the TDI observables can be computed \emph{numerically} from the LISA armlengths and plugged directly into the calculation of the likelihood, the essential ingredient of GW data analysis.
The mathematical counterparts of these qualitative statements are the formulation of a joint probability density for the phase measurements and laser noises, which is marginalized with respect to the latter to yield the likelihood used in data analysis; and the \emph{definition} of TDI observables as the null-space basis vectors of the design matrix that models the delayed appearance of the laser noises in the phase measurements.
We refer to these vectors as ``TDI-$\infty$'' observables, since they cancel laser noise for \emph{any} time dependence of the LISA armlengths, whereas ``first-generation'' TDI is limited to constant armlengths, ``second-generation'' TDI to linearly evolving armlengths with sufficiently small rates, and so on.

\section{Toy problem and discretized representation}
\label{sec:toy}

\begin{figure}
    \centering
    \includegraphics[width=0.9\columnwidth]{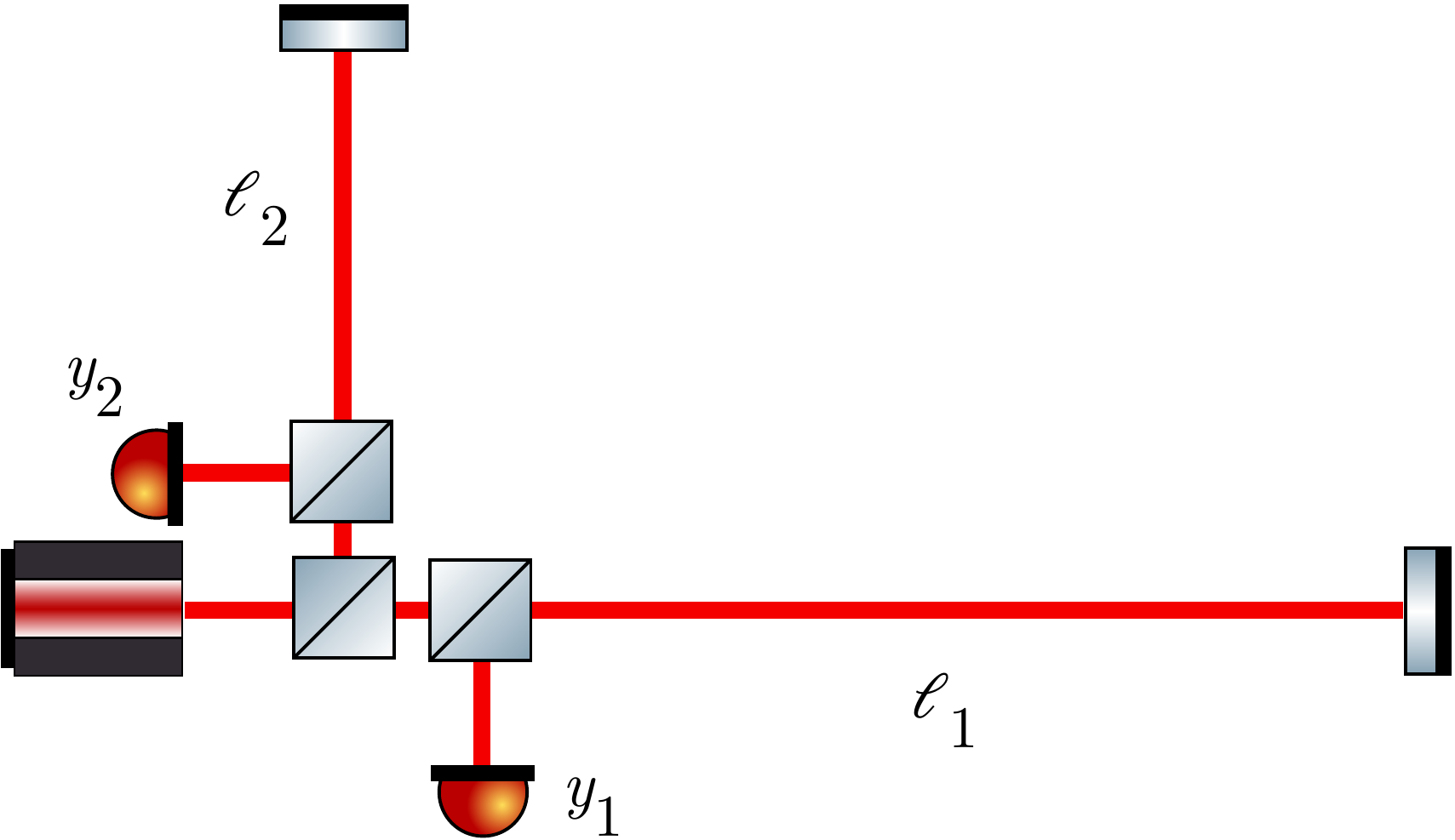}
    \caption{Setup of our toy model: a single laser source is propagated into arms with lengths $\ell_{1,2}(t)$ and reflected back toward the origin. The phases of the two beams are measured as $y_{1,2}(t)$, and are subject to the common laser noise $c(t)$, and to measurement noises $n_{1,2}(t)$.
    \label{fig:setup}}
\end{figure}

In this paper we describe our proposed scheme in the context of a representative toy model of the LISA measurements (see Fig.\ \ref{fig:setup}). We consider a single laser $c(t)$, propagated along arms 1 and 2 (with lengths that may be evolving with time), and reflected back by perfect mirrors; phase measurements $y_1(t)$ and $y_2(t)$ are performed at the origin, separately for each arm. Thus the measurement can be written as 
\begin{equation}
    \begin{aligned}
        y_1(t) &= c(t - \ell_1(t)) - c(t) + n_1(t) \qc
        \\
        y_2(t) &= c(t - \ell_2(t)) - c(t) + n_2(t) \
    \end{aligned}
    \label{eq:measurement}
\end{equation}
where $\ell_{1,2}(t)$ are the roundtrip flight times along the two arms for light pulses that are received at time $t$, and $n_{1,2}(t)$ represent measurement noises. In terms of one-way armlengths we have
$\ell_1(t) = L_{\overleftarrow{1}}(t) + L_{\overrightarrow{1}}(t - L_{\overleftarrow{1}}(t))$, with $L_{\overleftarrow{1}}$ the incoming flight time along arm 1, and $L_{\overrightarrow{1}}$ the outgoing flight time; the two will be different if the mirror is moving with respect to the origin. Crucially, we assume $c(t) \gg n_{1,2}(t)$. We do not model gravitational waves, but they would appear in both $y_1$ and $y_2$ with appropriate delays and geometric projections.

In practical measurements, all continuous time series will be sampled discretely with sufficiently high cadence, so in what follows we adopt the language and notation of linear algebra. Doing so is appropriate also for laser noise, under the assumption that interferometric signals are filtered so that the Nyquist criterion is satisfied by the sampling.

It is convenient to combine the two measurements $y_{1,2}$ and their noises $n_{1,2}$ into vectors $\mathbf{y}$ and $\mathbf{n}$, so we write
\begin{equation}
\label{eq:vectorized}
    \mathbf{y} = \mathsf{M} \, \mathbf{c} + \mathbf{n};
\end{equation}
here $\mathbf{c}$ is the (discretized) laser noise time series, and $\mathsf{M}$ is a \emph{design matrix} that models the delayed finite differences of Eq.\ \eqref{eq:measurement} by way of fractional-delay finite-impulse-response filters.
These very filters will be used in the post-processing of the LISA data to delay the interferometric measurements as required in TDI (see below).
Therefore the approximation that we make by writing Eq.\ \eqref{eq:vectorized} as a discrete equation is already implicitly accepted in standard usage.

If we assume (without loss of generality) that the laser noises are switched on instantaneously at time $t=0$, and that the delays $\ell_1$ and $\ell_2$ are constant multiples $2 \Delta t$ and $3 \Delta t$ of the basic sample cadence, the application of the design matrix would look like
\begin{equation}
\left(
    \begin{array}{c}
        y_1(t_0) \\
        y_2(t_0) \\
        y_1(t_1) \\
        y_2(t_1) \\
        y_1(t_2) \\
        y_2(t_2) \\
        y_1(t_3) \\
        y_2(t_3) \\
        y_1(t_4) \\
        y_2(t_4) \\
        \vdots
    \end{array}
\right)
=
\left(
    \begin{array}{rrrrrr}
        -1 &  0 &  0 & 0  & 0  & \cdots \\
        -1 &  0 &  0 & 0  & 0  & \cdots \\
         0 & -1 &  0 & 0  & 0  & \cdots \\
         0 & -1 &  0 & 0  & 0  & \cdots \\
         1 &  0 & -1 & 0  & 0  & \cdots \\
         0 &  0 & -1 & 0  & 0  & \cdots \\
         0 &  1 &  0 & -1 & 0  & \cdots \\
         1 &  0 &  0 & -1 & 0  & \cdots \\
         0 &  0 &  1 &  0 & -1 & \cdots \\
         0 &  1 &  0 &  0 & -1 & \cdots \\
         \vdots & \vdots & \vdots & \vdots & \vdots & \ddots
    \end{array}
\right)
\cdot
\left(
    \begin{array}{c}
        c(t_0) \\
        c(t_1) \\
        c(t_2) \\
        c(t_3) \\
        c(t_4) \\
        \vdots
    \end{array}
\right),
\end{equation}
where $t_k = k \Delta t$, and where we have interleaved $y_1$ and $y_2$ measurements.
If we obtain measurements at $n$ epochs, then $\mathbf{c}$ is an $n$-vector, $\mathbf{y}$ a $2n$-vector, and $\mathsf{L}$ a $2n \times n$ matrix.

Fractional delays would spread out the leftmost 1s into the appropriate filter masks. In this paper we shall use delay filters based on \emph{Lagrange interpolation}: that is, the $m$-point filter mask follows from approximating $f(\delta t)$, with $0 < \delta t < 1$, by evaluating the $(m - 1)$-order interpolating polynomial with nodes at $f(-m/2+1)$, $f(-m/2+2)$, \ldots, $f(-1)$, $f(0)$, $f(1)$, \ldots, $f(m/2)$. (In this illustration, for simplicity we have set the cadence $\Delta t$ equal to 1.) Filters with $\lfloor \delta t \rfloor \neq 0$ are obtained by first shifting the nodes by that integer part, then evaluating the interpolating polynomial at $\delta t - \lfloor \delta t \rfloor$. These filters have the property of maximal flatness in the frequency domain at $f = 0$; we always use them with even $m$, so that fractionally delayed quantities are continuous across $\delta t = 1/2$.
\begin{figure}
\includegraphics[width=\columnwidth]{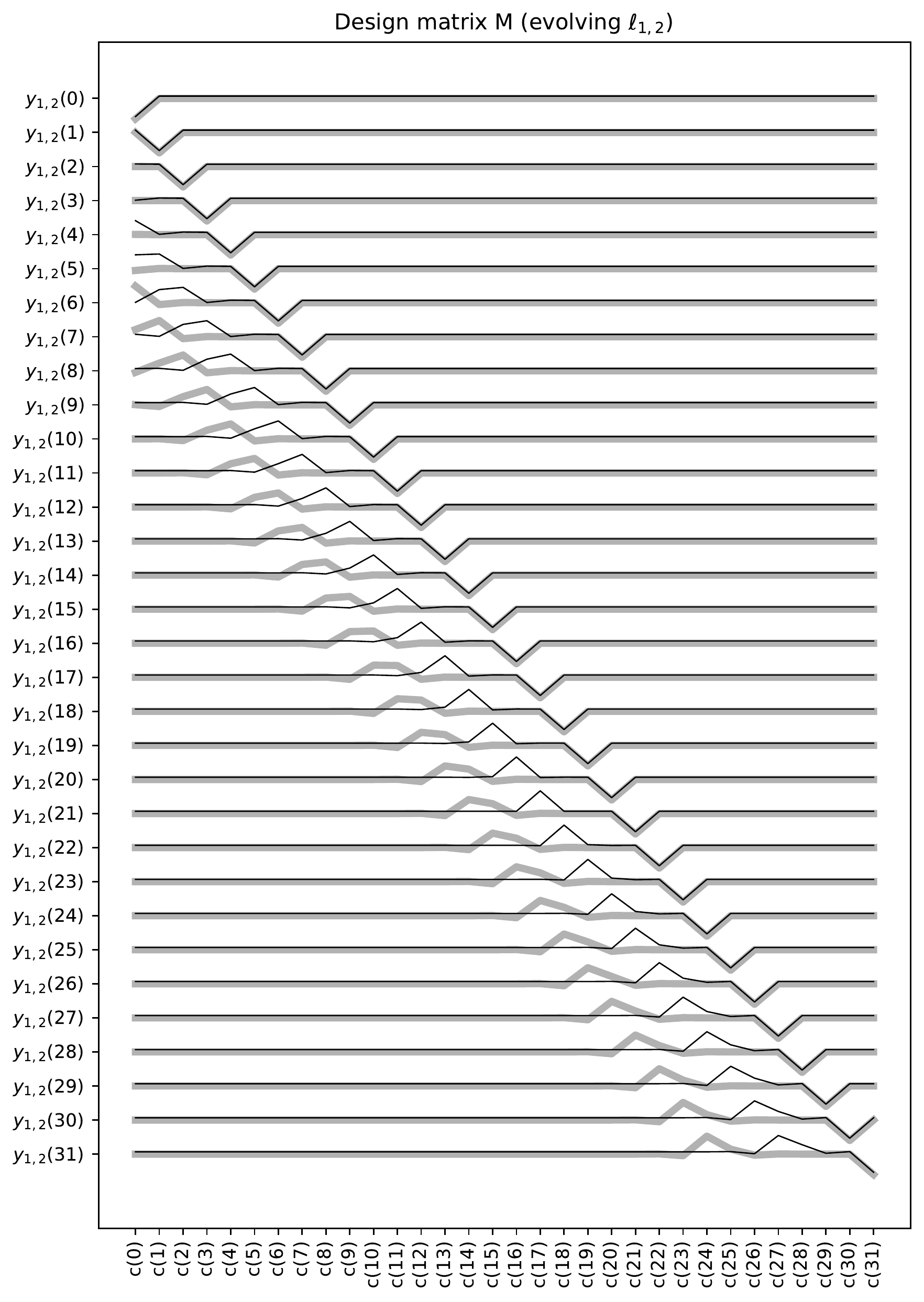}
\caption{Visualization of the design matrix $\mathsf{M}$ that maps the laser noise $\mathbf{c}$ to the phase measurements $\mathbf{y}_{1,2}$. The thick gray (thin black) lines plot the coefficients that multiply the $\mathbf{c}$ to yield the $\mathbf{y}_{1}$ ($\mathbf{y}_{2}$).
Here light-propagation delays are set as 
$\ell_1(t) = 6.2 + 0.02 t$ and $\ell_2(t) = 4.6 - 0.03 t$, and are implemented with six-point Lagrange-interpolation filters.
For simplicity, $c(t) = 0$ for $t < 0$.
\label{fig:mmat}}
\end{figure}

Figure \ref{fig:mmat} shows a graphical representation of the design matrix, this time for linearly evolving, non-integer delays implemented with fractional-delay masks of length $m = 6$.
Odd and even rows, corresponding to $y_1$ and $y_2$, are shown as thicker and lighter lines respectively.
The diagonal pattern of dips, common to both $y_1(t)$ and $y_2(t)$, corresponds to the ``direct'' $-c(t)$ terms.
The patterns below the diagonal correspond to delayed $c(t - \ell_{1,2}(t))$ terms, as realized by way of fractional-delay filter masks, and they are seen to shift with changing $\ell_{1,2}(t)$.

\section{Classical TDI}

In the classical TDI approach \cite{Tinto:1999yr,Armstrong_1999,Tinto:2002de,Tinto:2004,TintoLRR2014}, one derives laser-noise-free \emph{TDI observables} written as linear combinations of delayed measurements $y_{1,2}$. In our toy model there is one single such observable, which we may identify with the standard Michelson combination $M$ if $\ell_{1,2}$ are equal and constant:
\begin{equation}
\label{eq:tdim}
    M(t) = y_1(t) - y_2(t);
\end{equation}
with ``first-generation TDI'' $X$ if $\ell_{1,2}$ are \emph{unequal} and constant:
\begin{equation}
\label{eq:tdix}
\begin{aligned}
    X(t) = & \bigl(y_1(t) + y_2(t - \ell_1(t))\bigr) - \\
           & \bigl(y_2(t) + y_1(t - \ell_2(t))\bigr);
\end{aligned}
\end{equation}
and with ``second-generation TDI'' $X_1$ if $\ell_{1,2}$ are unequal and mildly evolving:
\begin{equation}
\label{eq:tdix1}
\begin{aligned}
    X_1(t) =
             & \bigl(y_1(t) + y_2(t_{,1}) + y_2(t_{,12}) + y_1(t_{,122})\bigr) - \\
             & \bigl(y_2(t) + y_1(t_{,2}) + y_1(t_{,21}) + y_2(t_{,211})\bigr),
\end{aligned}
\end{equation}
where $t_{,1} = t - \ell_1(t)$, $t_{,12} = t - \ell_1(t) - \ell_2(t - \ell_1(t))$, and so on. By inserting Eq.\ \eqref{eq:measurement} into Eqs.\ \eqref{eq:tdim}--\eqref{eq:tdix1} one can verify that the laser noises cancel in pairs. This is trivial for $M$. In each line of the equations for $X$ and $X_1$, the delayed-laser term of each measurement cancels the direct term on the next one; also, the direct terms of $y_1(t)$ and $y_2(t)$ cancel out, as do the delayed terms of the last measurements. This last cancellation is only approximate for $X_1$ when the delays are unequal and evolving: Taylor-expanding all laser noises, one sees that TDI cancels noise terms that are linear in $\dot{\ell}_{1,2}$, but not those of higher orders (terms that are $O(\dot{\ell}^2_{1,2})$, $O(\ddot{\ell}_{1,2})$, and so on).
Nevertheless, second-generation TDI is sufficient to reduce laser noise to levels compatible with the LISA requirements, as shown by experiments \cite{deVine:2010vf,Schwarze:2016fbv,Laporte:2017iuk,Cruz:2006js} and analytic and numerical studies \cite{Vallisneri:2004bn,Petiteau:2008zz,Bayle:2019}.

This procedure has a beautiful geometric formulation in terms of synthesized interferometric paths \cite{Vallisneri:2005}. It can also be formalized algebraically in terms of polynomial syzygies \cite{RajeshNayak:2004jzp}.
Last, it can be recast as an application of principal component analysis \cite{Romano2006,Baghi2020}---a formalism closely related to ours, and discussed further below.

Going back to our linear-algebraic notation, we represent a TDI observable evaluated at times $t_0$, $t_1$, \ldots as the vector
\begin{equation}
\mathbf{o} = \mathsf{T} \mathbf{y},
\end{equation}
with $\mathsf{T}$ an $n \times 2n$ matrix that encodes the delays of Eqs.\ \eqref{eq:tdim}--\eqref{eq:tdix1} by way of fractional-delay filters. Laser-noise cancellation then corresponds to
\begin{equation}
\label{eq:tm0}
\mathsf{T} \mathsf{M} \simeq 0,
\end{equation}
where the cancellation is exact for $M$ (or $X$) with constant and equal (unequal) $\ell_{1,2}$, and approximate but very accurate for $X_1$ with evolving and unequal $\ell_{1,2}$.

GW searches and source parameter estimation proceed from the evaluation of the likelihood of the data as a function of GW parameters $\theta$. In terms of the TDI vector $\mathbf{o}$, we obtain the likelihood by postulating that the measurement noises $n_{1,2}(t)$ are independent Gaussian processes with zero mean and covariance  $N(t',t'')$, and by equating the probability of observing the measurement residual $\Delta \mathbf{o} = \mathsf{T} (\mathbf{y} - \mathbf{y}_\mathrm{GW}(\theta))$ to the sampling probability of the noise, appropriately mapped from $\mathbf{y}$ to $\mathbf{o}$:
\begin{multline}
\label{eq:tlike}
\log p(\Delta \mathbf{o} = \mathsf{T} \Delta \mathbf{y}|\theta) = \\
= -\frac{1}{2} \Delta \mathbf{o}^\dagger (\mathsf{T} \mathsf{N} \mathsf{T}^\dagger)^{-1} \Delta \mathbf{o}
-\frac{1}{2} \log |2 \pi \mathsf{T} \mathsf{N} \mathsf{T}^\dagger|,
\end{multline}
where
\begin{equation}
    \mathsf{N}_{(ai)(bj)} = \bigl\langle y_a(t_i) y_b(t_j) \bigr\rangle = \delta_{ab} N(t_i, t_j).
\end{equation}

In the classical treatments of TDI, one would usually compute the spectral density $S_X(f)$ of the TDI observable as a function of acceleration and position noise in each element of LISA, and then write the log likelihood as 
\begin{equation}
    -2 \, \mathrm{Re} \! \int \frac{\Delta \tilde{X}^*(f)  \Delta \tilde{X}(f)}{S_X(f)} \, \mathrm{d} f;
\end{equation}
this equation is exactly equivalent to Eq.\ \eqref{eq:tlike}, where $\mathsf{T} \mathsf{N} \mathsf{T}^\dagger$ plays the role of $S_X(f)$ in the discretized time domain.

\section{Introducing TDI Infinity}

Instead of formulating the TDI observables algebraically or geometrically by matching direct and delayed noise terms, resulting in equations similar to \eqref{eq:tdim}--\eqref{eq:tdix1}, we take the approach of \emph{defining} the set of discretized TDI vectors by solving the matrix equation
\begin{equation}
\label{eq:tmeq}
\mathsf{T} \mathsf{M} = 0
\end{equation}
for $\mathsf{T}$ given the design matrix $\mathsf{M}$, which is determined by the LISA orbits and by the accurate times at which the $y_{1,2}$ are sampled. (See also Ref.\ \cite{2008CQGra..25f5005V}, which derives the TDI observables by solving Eq.\ \eqref{eq:tmeq} in the frequency domain, under the assumption that the armlengths are constant.)

The solution $\hat{\mathsf{T}}$ to Eq.\ \eqref{eq:tmeq}, which is unique up to affine transformations, provides a basis for the \emph{null space} of $\mathsf{M}^\dagger$: any vector $\hat{\mathbf{o}}$ within the null space solves the equation $\mathsf{M}^\dagger \hat{\mathbf{o}} = 0$. Correspondingly, each row $\hat{\mathbf{o}}_k$ of $\hat{\mathsf{T}}$ can be dotted into an observed vector $\mathbf{y}$ to generate a laser-noise--free observation $o_k$. We refer to the rows $\hat{\mathbf{o}}_k$ as \emph{TDI-$\infty$ observables}; by construction, they cancel laser noise for \emph{any} time dependence of the light-propagation delays. Given that $\mathsf{M}$ is a $2 n \times n$ matrix of rank $n$, we obtain $n$ linearly independent TDI-$\infty$ observables.

It should be clear from our theoretical development so far that $\hat{\mathsf{T}}$ can be used with Eq.\ \eqref{eq:tlike} to evaluate the TDI likelihood directly from the interferometric measurements $\mathbf{y}$, without the additional step of computing time-delayed combinations of measurements and GW templates.
Furthermore, the solution of Eq.\ \eqref{eq:tmeq} and the computation of the inverse covariance matrix $\mathsf{K}^{-1} \equiv (\hat{\mathsf{T}} \mathsf{N} \hat{\mathsf{T}}^\dagger)^{-1}$ can be performed \emph{offline}, before the repeated evaluation of the likelihood in a search or parameter-estimation scheme. The \emph{online} steps are the TDI-$\infty$ projection $\Delta \hat{\mathbf{o}} = \hat{\mathsf{T}} (\mathbf{y} - \mathbf{y}_\mathrm{GW}(\theta))$ and the kernel product $-\frac{1}{2} \Delta \hat{\mathbf{o}}^\dagger \mathsf{K}^{-1} \Delta  \hat{\mathbf{o}}$.

We further motivate our proposal by demonstrating that, in the LISA-appropriate limit of large laser noise, the TDI-$\infty$ likelihood is equivalent to the likelihood written from first principles for the $y$ measurements. That is, we can derive TDI-$\infty$ from a complete generative model of the LISA measurements, without need to model laser-noise subtraction explicitly.

Representing $c(t)$ as a Gaussian process with mean zero and covariance function $C(t',t'')$ (with  $\mathsf{C}_{ij} \equiv C(t_i, t_j)$), we write the likelihood of $\mathbf{c}$ and of the observed residuals $\Delta \mathbf{y} = \mathbf{y} - \mathbf{y}_\mathrm{GW}(\theta)$ as
\begin{equation}
     p(\Delta \mathbf{y},\mathbf{c}|\theta) = |2 \pi \mathsf{N}|^{-1/2} \, \mathrm{e}^{-\frac{1}{2} (\Delta \mathbf{y}(\theta) - \mathsf{M} \mathbf{c})^\dagger \mathsf{N}^{-1} (\Delta \mathbf{y}(\theta) - \mathsf{M} \mathbf{c})};
\end{equation}
integrating this likelihood with respect to $\mathbf{c}$, after multiplying by their prior $p(\mathbf{c}) = |2 \pi \mathsf{C}|^{-1/2} \, \mathrm{e}^{-\frac{1}{2} \mathbf{c}^\dagger \mathsf{C}^{-1} \mathbf{c}}$, yields the \emph{marginalized} log likelihood \cite{gpbook2006}
\begin{equation}
\label{eq:gplike}
\begin{aligned}
\log p(\Delta \mathbf{y}|\theta) = &-\frac{1}{2} \Delta \mathbf{y}^\dagger(\theta)
(\mathsf{N} + \mathsf{M} \mathsf{C} \mathsf{M}^\dagger)^{-1} \Delta \mathbf{y}(\theta) \\ &- \frac{1}{2} \log |2 \pi (\mathsf{N} + \mathsf{M} \mathsf{C} \mathsf{M}^\dagger)|.
\end{aligned}
\end{equation}
The marginalization can be seen as a probabilistic version of solving for the lasers, and then propagating the uncertainty of the solution to the remaining degrees of freedom. In Eq.\ \eqref{eq:gplike}, the augmentation of the covariance matrix $\mathsf{N}$ by $\mathsf{M} \mathsf{C} \mathsf{M}^\dagger$  has the effect of downweighting (or, in the limit $c(t) \gg n_{1,2}(t)$, completely projecting out) the linear combinations of the $\mathbf{y}$ in which the laser noises are dominant.

While Eq.\ \eqref{eq:gplike} could be used directly for GW applications, doing so carries the risk of losing numerical precision, possibly catastrophically.
The reason is that for LISA the $\mathbf{y}$ will always be strongly dominated by the laser noise $\mathbf{c}$; while the specific form of the covariance matrix will (in effect) select the $\mathbf{c}$-orthogonal components of the $\mathbf{y}$, that projection will involve the dangerous cancellation of very large numbers.

We can instead rely on Eqs.\ \eqref{eq:tlike} and \eqref{eq:tmeq}, which we show to be equivalent to Eq.\ \eqref{eq:gplike} in the limit of overwhelming laser noise.
To realize that limit, we take $\mathsf{C} = \sigma \mathsf{1}$ with $\sigma \rightarrow \infty$, and write the inverse Gaussian-process kernel of Eq.\ \eqref{eq:gplike} using the singular value decomposition (SVD) $\mathsf{M} = \mathsf{U} \mathsf{S} \mathsf{V}^\dagger$:
%
\begin{equation}
\begin{aligned}
(\mathsf{N} + \mathsf{M} \mathsf{C} \mathsf{M}^\dagger)^{-1} & = 
(\mathsf{N} + \sigma \mathsf{U} \mathsf{S} \mathsf{S}^\dagger \mathsf{U}^\dagger)^{-1} \\ &=
\mathsf{U} (\mathsf{U}^\dagger \mathsf{N} \mathsf{U} + \sigma \mathsf{S} \mathsf{S}^\dagger)^{-1} \mathsf{U}^\dagger,
\end{aligned}
\label{eq:step1}
\end{equation}
where the second equality follows by inserting factors $\mathsf{U}\mathsf{U}^\dagger = \mathsf{I}$ and shifting the $2n \times 2n$ orthogonal matrix $\mathsf{U}$ outside the inverse. We then refactor the second line of Eq.\ \eqref{eq:step1} as a block-matrix product, subdividing the columns of $\mathsf{U}$ as $(\mathsf{E}, \mathsf{F})$, where $\mathsf{E}$ spans the range of $\mathsf{M}$ and $\mathsf{F}^\dagger$ the null space of $\mathsf{M}^\dagger$:
\begin{equation}
\label{eq:kernelblock}
\bigl( \mathsf{E} \, \mathsf{F} \bigr)
\left( \begin{array}{ll}
    \mathsf{E}^\dagger \mathsf{N} \mathsf{E} + \sigma \mathsf{S} \mathsf{S}^\dagger & \mathsf{E}^\dagger \mathsf{N} \mathsf{F} \\
    \mathsf{F}^\dagger \mathsf{N} \mathsf{E} & \mathsf{F}^\dagger \mathsf{N} \mathsf{F}
\end{array} \right)^{\!\!-1}
\left(\begin{array}{c} \mathsf{E}^\dagger \\ \mathsf{F}^\dagger \end{array}\right).
\end{equation}
Using the block inverse formula, we find that all blocks are $O(\sigma^{-1})$ and disappear in the limit $\sigma \rightarrow \infty$, except for the bottom right block $\mathsf{F}^\dagger \mathsf{N} \mathsf{F}$ (the Schur complement). Thus
\begin{multline}
\label{eq:limit}
    \lim_{\sigma \rightarrow \infty} \log p(\Delta \mathbf{y}|\theta) = \\ -\frac{1}{2}
\Delta \mathbf{y}^\dagger(\theta) \mathsf{F} (\mathsf{F}^\dagger \mathsf{N} \mathsf{F})^{-1} \mathsf{F}^\dagger \Delta \mathbf{y}(\theta) -\frac{1}{2} \log |2 \pi \, \mathsf{F}^\dagger \mathsf{N} \mathsf{F}|.
\end{multline}
This limiting procedure is similar in spirit and mathematical detail to the marginalization over timing-model corrections in the time-domain analysis of pulsar-timing-array data \cite{2013MNRAS.428.1147V,2014PhRvD..90j4012V}; in that case, as here, the degrees of freedom with very large variance are effectively projected out of the data vector.

Now, $\mathsf{F}$ is a $2n \times n$ orthogonal matrix such that $\mathsf{M}^\dagger \mathsf{F} = \mathsf{F}^\dagger \mathsf{M} = 0$; given that the TDI-$\infty$ matrix $\hat{\mathsf{T}}$ is full rank and that $\hat{\mathsf{T}} \mathsf{M} = 0$, there must exist an invertible but not necessarily orthogonal matrix $\mathsf{A}$ such that $\mathsf{F}^\dagger = \mathsf{A} \mathsf{T}$. Inserting this representation in Eq.\ \eqref{eq:limit} reproduces Eq.\ \eqref{eq:tlike}, modulo an additive factor that does not depend on $\mathsf{N}$.

In addition to demonstrating the large-$\mathbf{c}$ equivalence of Eqs.\ \eqref{eq:tlike} and \eqref{eq:gplike}, this derivation suggests that the numerical instability of Eq.\ \eqref{eq:gplike} is resolved in Eq.\ \eqref{eq:tlike}, since the large components proportional to $\mathsf{C}$ drop out of Eq.\ \eqref{eq:kernelblock}, while the $\hat{\mathsf{T}}$ projection cancels out the large laser-noise contributions to the $\mathbf{y}$ measurements. The projection does require sufficient measurement precision and linearity, but no more so than the computation of the delayed combinations of classical TDI.

In Ref.\ \cite{Romano2006}, Romano and Woan identify the TDI observables with the small-eigenvalue eigenvectors of the $\mathbf{y}$ covariance matrix ($\mathsf{N} + \mathsf{M} \mathsf{C} \mathsf{M}^\dagger$ in our notation), and emphasize that its singular value decomposition factorizes the $\mathbf{y}$ likelihood into a TDI term (a \emph{sufficient statistic} for astrophysical inference), and a laser-dominated term (useful for laser-noise monitoring but not GW detection). They also recover the classical TDI expressions by analyzing the covariance matrix for integer-$\Delta t$ laser delays.
In the limit of large laser noise, Romano and Woan's approach is equivalent to the null-space formulation discussed here: indeed, Eqs.\ \eqref{eq:step1} and \eqref{eq:kernelblock} describe how the SVD of $\mathsf{M}$ induces the factorization of the marginalized likelihood.
Baghi and colleagues \cite{Baghi2020} perform the Romano--Woan eigenvector decomposition in the frequency domain, and work with the resulting $\mathbf{y}$ likelihood to simultaneously fit the GW source parameters, the LISA armlengths, and the components of the covariance matrix.

\section{The observables of TDI infinity}
\label{sec:observables}
\begin{figure*}
\includegraphics[width=2.3in]{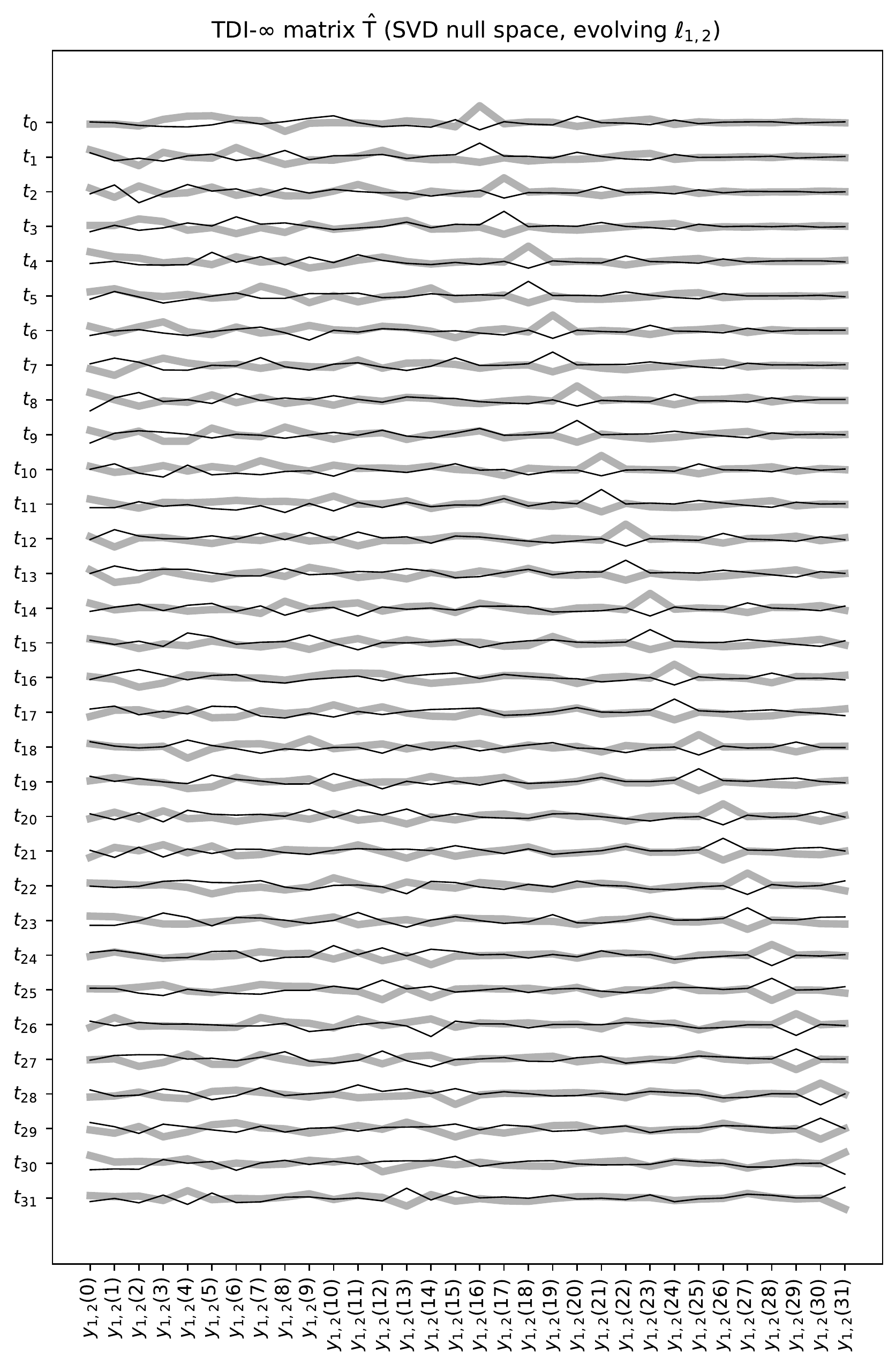}
\includegraphics[width=2.3in]{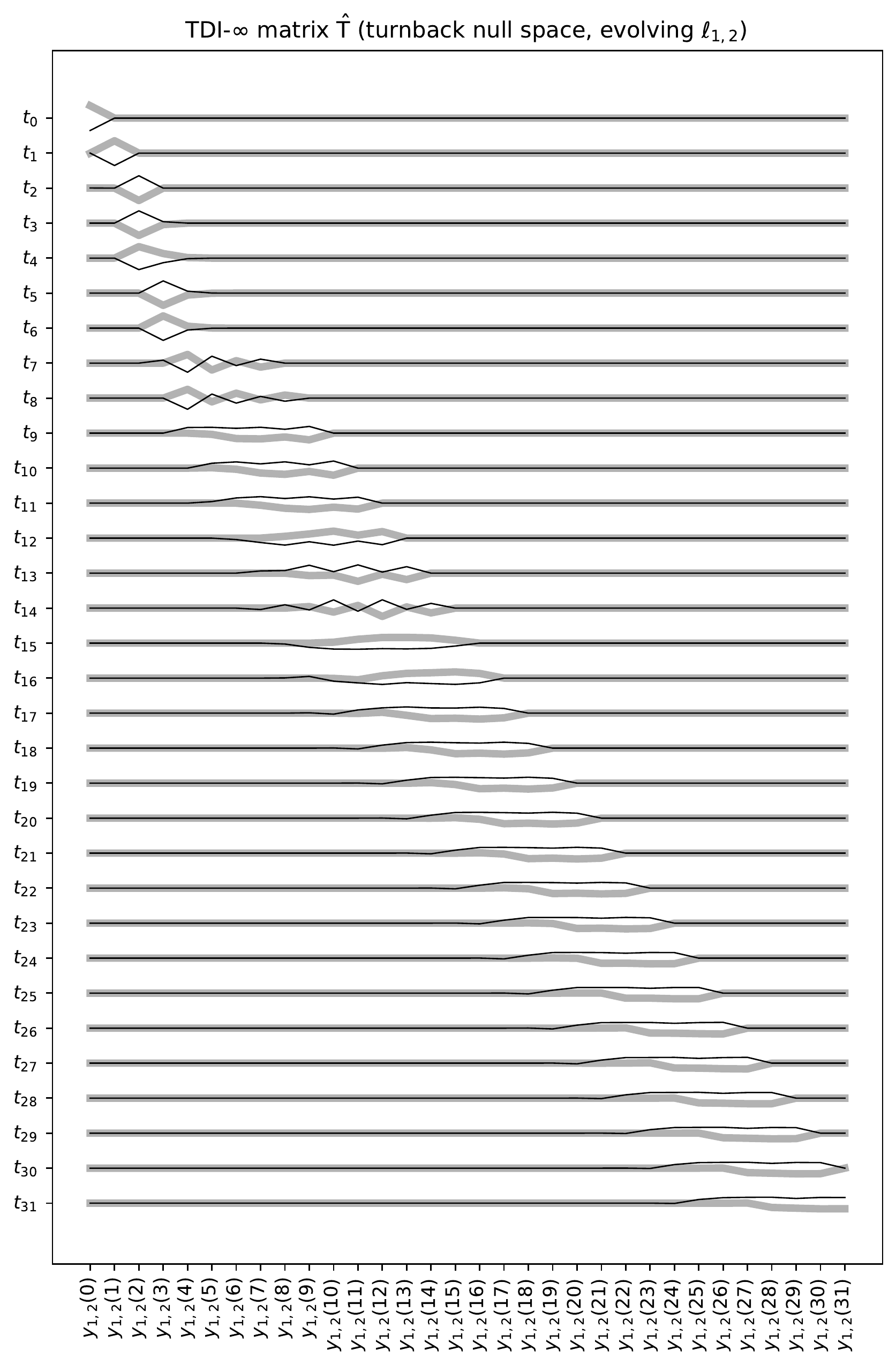}
\includegraphics[width=2.3in]{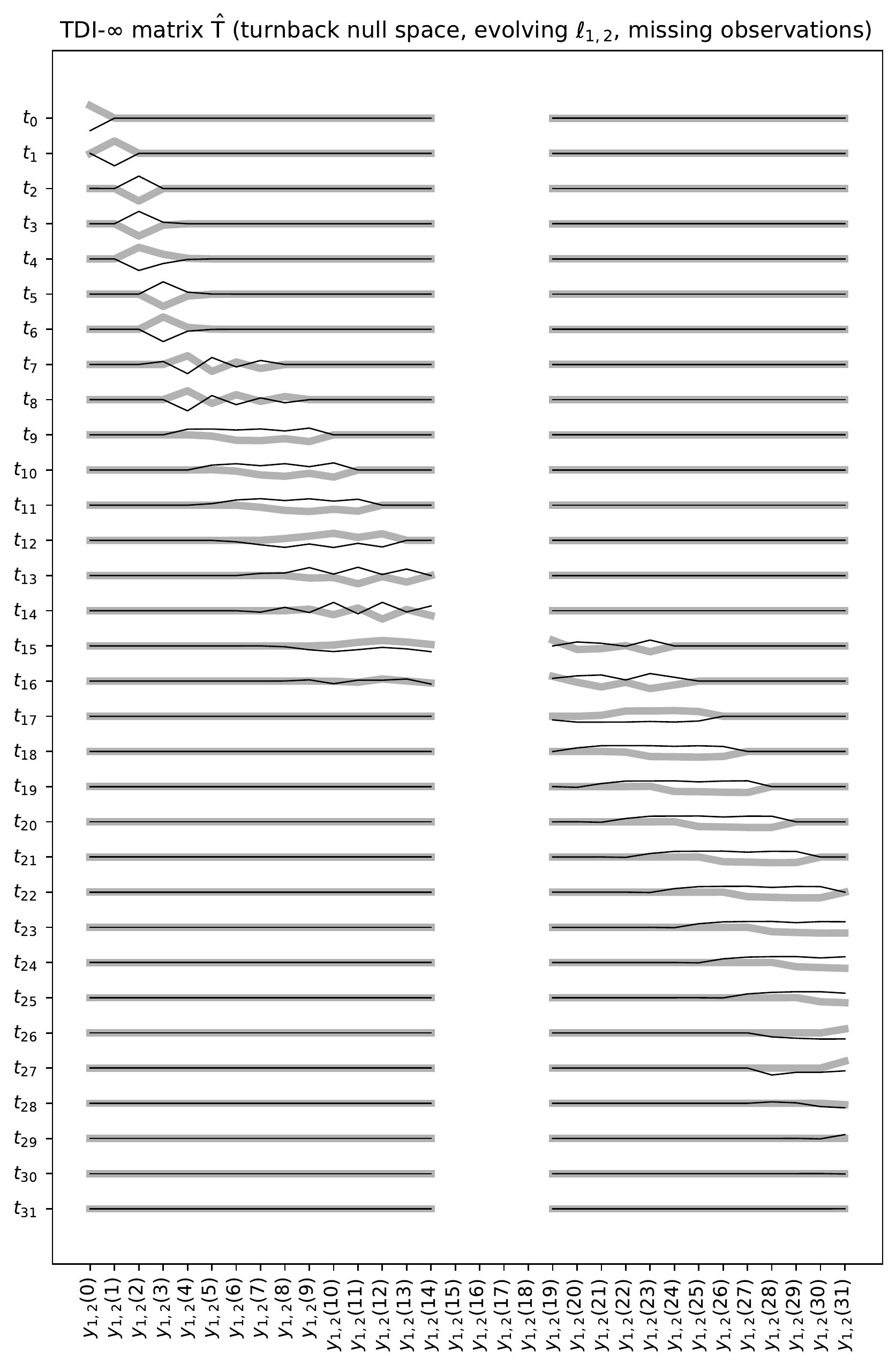}
\caption{
\textbf{Left}: TDI-$\infty$ vectors obtained by SVD from the design matrix of Fig.\ \ref{fig:mmat}. The thick gray (thin black) lines plot the coefficients that multiply the $\mathbf{y}_{1}$ ($\mathbf{y}_{2}$) to yield each laser-noise-free $t_k$ observation. The emergent diagonal structure on the right is an artifact of the specific SVD algorithm used here.
\textbf{Center}: TDI-$\infty$ vectors obtained by the turnback method \cite{topcu1979,kaneko1982,berry1985}, plotted with the same conventions.
These null-space basis vectors recover the banded diagonal structure of the design matrix (Fig.\ \ref{fig:mmat}) and the interpretation as time-local observables. 
The trivial structure at the top left is due to the simplifying assumption that $c(t) = 0$ for $t < 0$.
\textbf{Right}: TDI-$\infty$ vectors obtained by the turnback method when phase measurements $y_{1,2}(15)$ through $y_{1,2}(18)$ are missing. The last few TDI-$\infty$ vectors have analogous structure outside the range plotted here.
\label{fig:ymats}}
\end{figure*}

The standard linear-algebra approach to computing a basis for the null space of a matrix consists of factorizing it by SVD and then selecting the rows of the right factor that correspond to the null singular values.
These rows are orthogonal by construction, and in general they are dense across the matrix.
This means that TDI-$\infty$ vectors obtained from the SVD are \emph{nonlocal}: they span the length of the data, instead of being restricted to a few multiples of $\ell_{1,2}$ as the classical TDI observables.
The left panel of Fig.\ \ref{fig:ymats} shows such SVD vectors for the evolving-$\ell_{1,2}$ design matrix of Fig.\ \ref{fig:mmat}.
The particular SVD implementation used here (\texttt{dgesdd} from LAPACK \cite{laug}) results in some diagonal structure in the second half of the plotted timespan, but little regularity overall.

Dense TDI-$\infty$ vectors do not necessarily lead to loss of precision, but they certainly have other disadvantages. They obfuscate the time dependence of GW signals and instrument noise; they make it hard to analyze data in chunks; and they guarantee that the offline and online phases of likelihood evaluation have maximum computational complexities $O(n^3)$ and $O(n^2)$ (where $2 n$ is the length of the vector $\mathbf{y}$).

Fortunately, applied mathematicians have developed algorithms that generate \emph{banded} basis matrices for the null space of sparse banded matrices such as $\mathsf{M}^\dagger$. 
One such algorithm is the \emph{turnback} method, originally suggested by Topçu \cite{topcu1979} in the context of the matrix force method for linear elastic analysis, and further refined in Refs.\ \cite{kaneko1982,berry1985}. The turnback method begins with the standard LU decomposition \cite[e.g.][]{golub1996} of the sparse matrix, followed by a number of triangular factorizations of its submatrices. The resulting basis vectors are not orthogonal, but they are concentrated along the diagonal.

We experimented with the method using an implementation kindly provided by Thuy Van Dang and Keck Voon Ling \cite{dang2017}. For integer delays, turnback basis vectors reproduce exactly the observable $X$ of first-generation TDI (Eq.\ \eqref{eq:tdix}); for constant fractional delays, they have the same bandwidth as $X$ (the larger of the $\ell_{1}$ and $\ell_{2}$ plus half of the filter delay width), but with smoother structure; for evolving delays, as shown in the center panel of Fig.\ \ref{fig:ymats}, they have bandwidth comparable to $X$ rather than to $X_1$ (which has extent $\gtrsim 3 \times \ell_{1,2}$), again with smoother structure.
These statements discount the first few vectors at the top of the center panel, which have trivial structure because $c(t) = 0$ for $t < t_0$, so delayed laser-noise terms do not affect the $y_{1,2}$ until $t_7$.

We note that in a realistic data-reduction scenario the bandwidth is likely to be dominated by the length $m$ of the fractional-delay filters, which will span several LISA armlengths to achieve the required interpolation accuracy.
While the offline phase of likelihood evaluation has again complexity $O(n^3)$, as required by the turnback algorithm, the banded structure of $\mathsf{T}$ and therefore $\mathsf{T}^\dagger \mathsf{N} \mathsf{T}$ may allow the optimization of the online phase to complexities lower than $O(n^2)$.

In the right panel of Fig.\ \ref{fig:ymats} we demonstrate that the TDI-$\infty$ approach automatically takes data gaps into account. Here we have modeled missing phase measurements at four epochs by removing eight rows of the $2 n \times n$ design matrix. The solution of Eq.\ \eqref{eq:tmeq} with the turnback method yields $(n - 8)$ TDI-$\infty$ vectors that combine the measurements around the gap, shown as the blank vertical band in the plot. Notably, the $\ell_{1,2}(t)$ used here allow for two vectors that bridge the disruption. Near the gap, the bandwidth of the observables increases by $\sim 50\%$. By contrast, in this example the second-generation TDI observable $X_1$ would be unavailable for more than 20 epochs, since it requires phase observations spanning $\gtrsim 3 \times \ell_{1,2}$. This advantage washes out for longer gaps.

\section{Discussion}

By way of a toy model of LISA interferometry, we have offered a proof of principle that the LISA GW data analysis can be formulated and performed directly in terms of the phase measurements, without recourse to the analytical observables of classical TDI.
This approach leads to the numerically defined observables of TDI-$\infty$, which cancel laser noise for any time dependence of the armlengths, and which can be conveniently time-localized to bandwidths comparable to or smaller than those of second-generation TDI.
The scheme has several additional advantages:
\begin{itemize}
\item There is no need to select a set of analytical TDI observables, model their power-spectral densities, and track their data quality;
\item GW theoretical templates can be computed directly for the simpler phase measurements rather than the more complicated TDI observables, or even for the basic GW strain polarizations, and then projected to the phase measurements;
\item Measurement gaps are handled automatically and gracefully, including the shift between one, two, and three independent combinations when four, five, and six LISA laser links are available;
\item The link to the computation of matrix null-space bases, a linear-algebra problem with many practical applications, raises the possibility of adopting new sophisticated algorithms \cite[e.g.][]{eberly2004,coakley2011,kressner2017}, including parallelized or streaming variants suited to GPUs.
\end{itemize}

While our toy model is extremely idealized and therefore limited,  we believe these advantages warrant a detailed investigation of the numerical implementation of TDI-$\infty$ and of its implications for the LISA system, which we leave for future work.

\acknowledgments
MV is thankful to John Baker for posing the challenge of distributing the LISA data as the set of the ``$\mathbf{y}$'', and to Alvin Chua for helpful interactions and conversations.
MV is especially grateful to CEA/IPhT and APC for hospitality during a 2019 sabbatical that led to the inception of this work.
MV was supported by the JPL RTD program, and JBB by a NASA postdoctoral fellowship administered by USRA.
Part of this research was carried out at the Jet Propulsion Laboratory, California Institute of Technology, under a contract with the National Aeronautics and Space Administration (80NM0018D0004).
Copyright 2020. All rights reserved.

\bibliographystyle{apsrev4-1}
\bibliography{references}

\begin{thebibliography}{32}%
\makeatletter
\providecommand \@ifxundefined [1]{%
 \@ifx{#1\undefined}
}%
\providecommand \@ifnum [1]{%
 \ifnum #1\expandafter \@firstoftwo
 \else \expandafter \@secondoftwo
 \fi
}%
\providecommand \@ifx [1]{%
 \ifx #1\expandafter \@firstoftwo
 \else \expandafter \@secondoftwo
 \fi
}%
\providecommand \natexlab [1]{#1}%
\providecommand \enquote  [1]{``#1''}%
\providecommand \bibnamefont  [1]{#1}%
\providecommand \bibfnamefont [1]{#1}%
\providecommand \citenamefont [1]{#1}%
\providecommand \href@noop [0]{\@secondoftwo}%
\providecommand \href [0]{\begingroup \@sanitize@url \@href}%
\providecommand \@href[1]{\@@startlink{#1}\@@href}%
\providecommand \@@href[1]{\endgroup#1\@@endlink}%
\providecommand \@sanitize@url [0]{\catcode `\\12\catcode `\$12\catcode
  `\&12\catcode `\#12\catcode `\^12\catcode `\_12\catcode `\%12\relax}%
\providecommand \@@startlink[1]{}%
\providecommand \@@endlink[0]{}%
\providecommand \url  [0]{\begingroup\@sanitize@url \@url }%
\providecommand \@url [1]{\endgroup\@href {#1}{\urlprefix }}%
\providecommand \urlprefix  [0]{URL }%
\providecommand \Eprint [0]{\href }%
\providecommand \doibase [0]{http://dx.doi.org/}%
\providecommand \selectlanguage [0]{\@gobble}%
\providecommand \bibinfo  [0]{\@secondoftwo}%
\providecommand \bibfield  [0]{\@secondoftwo}%
\providecommand \translation [1]{[#1]}%
\providecommand \BibitemOpen [0]{}%
\providecommand \bibitemStop [0]{}%
\providecommand \bibitemNoStop [0]{.\EOS\space}%
\providecommand \EOS [0]{\spacefactor3000\relax}%
\providecommand \BibitemShut  [1]{\csname bibitem#1\endcsname}%
\let\auto@bib@innerbib\@empty
\bibitem [{\citenamefont {{Amaro-Seoane}}\ \emph {et~al.}(2017)\citenamefont
  {{Amaro-Seoane}} \emph {et~al.}}]{lisa2017}%
  \BibitemOpen
  \bibfield  {author} {\bibinfo {author} {\bibfnamefont {P.}~\bibnamefont
  {{Amaro-Seoane}}} \emph {et~al.},\ }\href@noop {} {\bibfield  {journal}
  {\bibinfo  {journal} {LISA Proposal for ESA Cosmic Visions L3}\ } (\bibinfo
  {year} {2017})},\ \Eprint {http://arxiv.org/abs/1702.00786} {arXiv:1702.00786
  [astro-ph.IM]} \BibitemShut {NoStop}%
\bibitem [{\citenamefont {Tinto}\ and\ \citenamefont
  {Armstrong}(1999)}]{Tinto:1999yr}%
  \BibitemOpen
  \bibfield  {author} {\bibinfo {author} {\bibfnamefont {M.}~\bibnamefont
  {Tinto}}\ and\ \bibinfo {author} {\bibfnamefont {J.~W.}\ \bibnamefont
  {Armstrong}},\ }\href {\doibase 10.1103/PhysRevD.59.102003} {\bibfield
  {journal} {\bibinfo  {journal} {Phys. Rev. D}\ }\textbf {\bibinfo {volume}
  {59}},\ \bibinfo {pages} {102003} (\bibinfo {year} {1999})}\BibitemShut
  {NoStop}%
\bibitem [{\citenamefont {Armstrong}\ \emph {et~al.}(1999)\citenamefont
  {Armstrong}, \citenamefont {Estabrook},\ and\ \citenamefont
  {Tinto}}]{Armstrong_1999}%
  \BibitemOpen
  \bibfield  {author} {\bibinfo {author} {\bibfnamefont {J.~W.}\ \bibnamefont
  {Armstrong}}, \bibinfo {author} {\bibfnamefont {F.~B.}\ \bibnamefont
  {Estabrook}}, \ and\ \bibinfo {author} {\bibfnamefont {M.}~\bibnamefont
  {Tinto}},\ }\href {\doibase 10.1086/308110} {\bibfield  {journal} {\bibinfo
  {journal} {Astrophys. J.}\ }\textbf {\bibinfo {volume} {527}},\ \bibinfo
  {pages} {814} (\bibinfo {year} {1999})}\BibitemShut {NoStop}%
\bibitem [{\citenamefont {Tinto}\ \emph {et~al.}(2002)\citenamefont {Tinto},
  \citenamefont {Estabrook},\ and\ \citenamefont {Armstrong}}]{Tinto:2002de}%
  \BibitemOpen
  \bibfield  {author} {\bibinfo {author} {\bibfnamefont {M.}~\bibnamefont
  {Tinto}}, \bibinfo {author} {\bibfnamefont {F.~B.}\ \bibnamefont
  {Estabrook}}, \ and\ \bibinfo {author} {\bibfnamefont {J.~W.}\ \bibnamefont
  {Armstrong}},\ }\href {\doibase 10.1103/PhysRevD.65.082003} {\bibfield
  {journal} {\bibinfo  {journal} {Phys. Rev. D}\ }\textbf {\bibinfo {volume}
  {65}},\ \bibinfo {pages} {082003} (\bibinfo {year} {2002})}\BibitemShut
  {NoStop}%
\bibitem [{\citenamefont {Tinto}\ \emph {et~al.}(2004)\citenamefont {Tinto},
  \citenamefont {Estabrook},\ and\ \citenamefont {Armstrong}}]{Tinto:2004}%
  \BibitemOpen
  \bibfield  {author} {\bibinfo {author} {\bibfnamefont {M.}~\bibnamefont
  {Tinto}}, \bibinfo {author} {\bibfnamefont {F.~B.}\ \bibnamefont
  {Estabrook}}, \ and\ \bibinfo {author} {\bibfnamefont {J.~W.}\ \bibnamefont
  {Armstrong}},\ }\href {\doibase 10.1103/PhysRevD.69.082001} {\bibfield
  {journal} {\bibinfo  {journal} {Phys. Rev. D}\ }\textbf {\bibinfo {volume}
  {69}},\ \bibinfo {pages} {082001} (\bibinfo {year} {2004})}\BibitemShut
  {NoStop}%
\bibitem [{\citenamefont {{Tinto}}\ and\ \citenamefont
  {{Dhurandhar}}(2014)}]{TintoLRR2014}%
  \BibitemOpen
  \bibfield  {author} {\bibinfo {author} {\bibfnamefont {M.}~\bibnamefont
  {{Tinto}}}\ and\ \bibinfo {author} {\bibfnamefont {S.~V.}\ \bibnamefont
  {{Dhurandhar}}},\ }\href {\doibase 10.12942/lrr-2014-6} {\bibfield  {journal}
  {\bibinfo  {journal} {Liv. Rev. Relativity}\ }\textbf {\bibinfo {volume}
  {17}},\ \bibinfo {eid} {6} (\bibinfo {year} {2014})}\BibitemShut {NoStop}%
\bibitem [{\citenamefont {{Baghi}}\ \emph {et~al.}(2019)\citenamefont
  {{Baghi}}, \citenamefont {{Thorpe}}, \citenamefont {{Slutsky}}, \citenamefont
  {{Baker}}, \citenamefont {{Canton}}, \citenamefont {{Korsakova}},\ and\
  \citenamefont {{Karnesis}}}]{2019PhRvD.100b2003B}%
  \BibitemOpen
  \bibfield  {author} {\bibinfo {author} {\bibfnamefont {Q.}~\bibnamefont
  {{Baghi}}}, \bibinfo {author} {\bibfnamefont {J.~I.}\ \bibnamefont
  {{Thorpe}}}, \bibinfo {author} {\bibfnamefont {J.}~\bibnamefont {{Slutsky}}},
  \bibinfo {author} {\bibfnamefont {J.}~\bibnamefont {{Baker}}}, \bibinfo
  {author} {\bibfnamefont {T.~D.}\ \bibnamefont {{Canton}}}, \bibinfo {author}
  {\bibfnamefont {N.}~\bibnamefont {{Korsakova}}}, \ and\ \bibinfo {author}
  {\bibfnamefont {N.}~\bibnamefont {{Karnesis}}},\ }\href {\doibase
  10.1103/PhysRevD.100.022003} {\bibfield  {journal} {\bibinfo  {journal}
  {\prd}\ }\textbf {\bibinfo {volume} {100}},\ \bibinfo {eid} {022003}
  (\bibinfo {year} {2019})}\BibitemShut {NoStop}%
\bibitem [{\citenamefont {{Hartwig}}\ and\ \citenamefont
  {{Bayle}}(2020)}]{2020arXiv200502430H}%
  \BibitemOpen
  \bibfield  {author} {\bibinfo {author} {\bibfnamefont {O.}~\bibnamefont
  {{Hartwig}}}\ and\ \bibinfo {author} {\bibfnamefont {J.-B.}\ \bibnamefont
  {{Bayle}}},\ }\href@noop {} {\bibfield  {journal} {\bibinfo  {journal} {arXiv
  e-prints}\ } (\bibinfo {year} {2020})},\ \Eprint
  {http://arxiv.org/abs/2005.02430} {arXiv:2005.02430 [astro-ph.IM]}
  \BibitemShut {NoStop}%
\bibitem [{\citenamefont {Bayle}\ \emph {et~al.}(2019)\citenamefont {Bayle},
  \citenamefont {Lilley}, \citenamefont {Petiteau},\ and\ \citenamefont
  {Halloin}}]{Bayle:2019}%
  \BibitemOpen
  \bibfield  {author} {\bibinfo {author} {\bibfnamefont {J.-B.}\ \bibnamefont
  {Bayle}}, \bibinfo {author} {\bibfnamefont {M.}~\bibnamefont {Lilley}},
  \bibinfo {author} {\bibfnamefont {A.}~\bibnamefont {Petiteau}}, \ and\
  \bibinfo {author} {\bibfnamefont {H.}~\bibnamefont {Halloin}},\ }\href
  {\doibase 10.1103/PhysRevD.99.084023} {\bibfield  {journal} {\bibinfo
  {journal} {Phys. Rev. D}\ }\textbf {\bibinfo {volume} {99}},\ \bibinfo
  {pages} {084023} (\bibinfo {year} {2019})}\BibitemShut {NoStop}%
\bibitem [{\citenamefont {de~Vine}\ \emph {et~al.}(2010)\citenamefont
  {de~Vine}, \citenamefont {Ware}, \citenamefont {McKenzie}, \citenamefont
  {Spero}, \citenamefont {Klipstein},\ and\ \citenamefont
  {Shaddock}}]{deVine:2010vf}%
  \BibitemOpen
  \bibfield  {author} {\bibinfo {author} {\bibfnamefont {G.}~\bibnamefont
  {de~Vine}}, \bibinfo {author} {\bibfnamefont {B.}~\bibnamefont {Ware}},
  \bibinfo {author} {\bibfnamefont {K.}~\bibnamefont {McKenzie}}, \bibinfo
  {author} {\bibfnamefont {R.~E.}\ \bibnamefont {Spero}}, \bibinfo {author}
  {\bibfnamefont {W.~M.}\ \bibnamefont {Klipstein}}, \ and\ \bibinfo {author}
  {\bibfnamefont {D.~A.}\ \bibnamefont {Shaddock}},\ }\href {\doibase
  10.1103/PhysRevLett.104.211103} {\bibfield  {journal} {\bibinfo  {journal}
  {Phys. Rev. Lett.}\ }\textbf {\bibinfo {volume} {104}},\ \bibinfo {pages}
  {211103} (\bibinfo {year} {2010})}\BibitemShut {NoStop}%
\bibitem [{\citenamefont {Schwarze}\ \emph {et~al.}(2016)\citenamefont
  {Schwarze}, \citenamefont {Fernández~Barranco}, \citenamefont {Penkert},
  \citenamefont {Gerberding}, \citenamefont {Heinzel},\ and\ \citenamefont
  {Danzmann}}]{Schwarze:2016fbv}%
  \BibitemOpen
  \bibfield  {author} {\bibinfo {author} {\bibfnamefont {T.}~\bibnamefont
  {Schwarze}}, \bibinfo {author} {\bibfnamefont {G.}~\bibnamefont
  {Fernández~Barranco}}, \bibinfo {author} {\bibfnamefont {D.}~\bibnamefont
  {Penkert}}, \bibinfo {author} {\bibfnamefont {O.}~\bibnamefont {Gerberding}},
  \bibinfo {author} {\bibfnamefont {G.}~\bibnamefont {Heinzel}}, \ and\
  \bibinfo {author} {\bibfnamefont {K.}~\bibnamefont {Danzmann}},\ }\href
  {\doibase 10.1088/1742-6596/716/1/012004} {\bibfield  {journal} {\bibinfo
  {journal} {J. Phys. Conf. Ser.}\ }\textbf {\bibinfo {volume} {716}},\
  \bibinfo {pages} {012004} (\bibinfo {year} {2016})}\BibitemShut {NoStop}%
\bibitem [{\citenamefont {Laporte}\ \emph {et~al.}(2017)\citenamefont
  {Laporte}, \citenamefont {Halloin}, \citenamefont {Bréelle}, \citenamefont
  {Buy}, \citenamefont {Grüning},\ and\ \citenamefont
  {Prat}}]{Laporte:2017iuk}%
  \BibitemOpen
  \bibfield  {author} {\bibinfo {author} {\bibfnamefont {M.}~\bibnamefont
  {Laporte}}, \bibinfo {author} {\bibfnamefont {H.}~\bibnamefont {Halloin}},
  \bibinfo {author} {\bibfnamefont {E.}~\bibnamefont {Bréelle}}, \bibinfo
  {author} {\bibfnamefont {C.}~\bibnamefont {Buy}}, \bibinfo {author}
  {\bibfnamefont {P.}~\bibnamefont {Grüning}}, \ and\ \bibinfo {author}
  {\bibfnamefont {P.}~\bibnamefont {Prat}},\ }\href {\doibase
  10.1088/1742-6596/840/1/012014} {\bibfield  {journal} {\bibinfo  {journal}
  {J. Phys. Conf. Ser.}\ }\textbf {\bibinfo {volume} {840}},\ \bibinfo {pages}
  {012014} (\bibinfo {year} {2017})}\BibitemShut {NoStop}%
\bibitem [{\citenamefont {Cruz}\ \emph {et~al.}(2006)\citenamefont {Cruz},
  \citenamefont {Thorpe}, \citenamefont {Hartman},\ and\ \citenamefont
  {Mueller}}]{Cruz:2006js}%
  \BibitemOpen
  \bibfield  {author} {\bibinfo {author} {\bibfnamefont {R.}~\bibnamefont
  {Cruz}}, \bibinfo {author} {\bibfnamefont {J.}~\bibnamefont {Thorpe}},
  \bibinfo {author} {\bibfnamefont {M.}~\bibnamefont {Hartman}}, \ and\
  \bibinfo {author} {\bibfnamefont {G.}~\bibnamefont {Mueller}},\ }\href
  {\doibase 10.1063/1.2405062} {\bibfield  {journal} {\bibinfo  {journal} {AIP
  Conf. Proc.}\ }\textbf {\bibinfo {volume} {873}},\ \bibinfo {pages} {319}
  (\bibinfo {year} {2006})}\BibitemShut {NoStop}%
\bibitem [{\citenamefont {Vallisneri}(2005{\natexlab{a}})}]{Vallisneri:2004bn}%
  \BibitemOpen
  \bibfield  {author} {\bibinfo {author} {\bibfnamefont {M.}~\bibnamefont
  {Vallisneri}},\ }\href {\doibase 10.1103/PhysRevD.71.022001} {\bibfield
  {journal} {\bibinfo  {journal} {Phys. Rev. D}\ }\textbf {\bibinfo {volume}
  {71}},\ \bibinfo {pages} {022001} (\bibinfo {year}
  {2005}{\natexlab{a}})}\BibitemShut {NoStop}%
\bibitem [{\citenamefont {Petiteau}\ \emph {et~al.}(2008)\citenamefont
  {Petiteau}, \citenamefont {Auger}, \citenamefont {Halloin}, \citenamefont
  {Jeannin}, \citenamefont {Plagnol}, \citenamefont {Pireaux}, \citenamefont
  {Regimbau},\ and\ \citenamefont {Vinet}}]{Petiteau:2008zz}%
  \BibitemOpen
  \bibfield  {author} {\bibinfo {author} {\bibfnamefont {A.}~\bibnamefont
  {Petiteau}}, \bibinfo {author} {\bibfnamefont {G.}~\bibnamefont {Auger}},
  \bibinfo {author} {\bibfnamefont {H.}~\bibnamefont {Halloin}}, \bibinfo
  {author} {\bibfnamefont {O.}~\bibnamefont {Jeannin}}, \bibinfo {author}
  {\bibfnamefont {E.}~\bibnamefont {Plagnol}}, \bibinfo {author} {\bibfnamefont
  {S.}~\bibnamefont {Pireaux}}, \bibinfo {author} {\bibfnamefont
  {T.}~\bibnamefont {Regimbau}}, \ and\ \bibinfo {author} {\bibfnamefont
  {J.-Y.}\ \bibnamefont {Vinet}},\ }\href {\doibase 10.1103/PhysRevD.77.023002}
  {\bibfield  {journal} {\bibinfo  {journal} {Phys. Rev. D}\ }\textbf {\bibinfo
  {volume} {77}},\ \bibinfo {pages} {023002} (\bibinfo {year}
  {2008})}\BibitemShut {NoStop}%
\bibitem [{\citenamefont {Vallisneri}(2005{\natexlab{b}})}]{Vallisneri:2005}%
  \BibitemOpen
  \bibfield  {author} {\bibinfo {author} {\bibfnamefont {M.}~\bibnamefont
  {Vallisneri}},\ }\href {\doibase 10.1103/PhysRevD.76.109903,
  10.1103/PhysRevD.72.042003} {\bibfield  {journal} {\bibinfo  {journal} {Phys.
  Rev. D}\ }\textbf {\bibinfo {volume} {72}},\ \bibinfo {pages} {042003}
  (\bibinfo {year} {2005}{\natexlab{b}})},\ \bibinfo {note} {[Erratum: Phys.
  Rev. D 76, 109903 (2007)]}\BibitemShut {NoStop}%
\bibitem [{\citenamefont {Rajesh~Nayak}\ and\ \citenamefont
  {Vinet}(2004)}]{RajeshNayak:2004jzp}%
  \BibitemOpen
  \bibfield  {author} {\bibinfo {author} {\bibfnamefont {K.}~\bibnamefont
  {Rajesh~Nayak}}\ and\ \bibinfo {author} {\bibfnamefont {J.}~\bibnamefont
  {Vinet}},\ }\href@noop {} {\bibfield  {journal} {\bibinfo  {journal} {Phys.
  Rev. D}\ }\textbf {\bibinfo {volume} {70}} (\bibinfo {year}
  {2004})}\BibitemShut {NoStop}%
\bibitem [{\citenamefont {{Romano}}\ and\ \citenamefont
  {{Woan}}(2006)}]{Romano2006}%
  \BibitemOpen
  \bibfield  {author} {\bibinfo {author} {\bibfnamefont {J.~D.}\ \bibnamefont
  {{Romano}}}\ and\ \bibinfo {author} {\bibfnamefont {G.}~\bibnamefont
  {{Woan}}},\ }\href {\doibase 10.1103/PhysRevD.73.102001} {\bibfield
  {journal} {\bibinfo  {journal} {\prd}\ }\textbf {\bibinfo {volume} {73}},\
  \bibinfo {eid} {102001} (\bibinfo {year} {2006})}\BibitemShut {NoStop}%
\bibitem [{\citenamefont {Baghi}\ \emph {et~al.}(2020)\citenamefont {Baghi},
  \citenamefont {Thorpe}, \citenamefont {Slutsky},\ and\ \citenamefont
  {Baker}}]{Baghi2020}%
  \BibitemOpen
  \bibfield  {author} {\bibinfo {author} {\bibfnamefont {Q.}~\bibnamefont
  {Baghi}}, \bibinfo {author} {\bibfnamefont {J.~I.}\ \bibnamefont {Thorpe}},
  \bibinfo {author} {\bibfnamefont {J.}~\bibnamefont {Slutsky}}, \ and\
  \bibinfo {author} {\bibfnamefont {J.}~\bibnamefont {Baker}},\ }\href@noop {}
  {\bibfield  {journal} {\bibinfo  {journal} {in preparation}\ } (\bibinfo
  {year} {2020})}\BibitemShut {NoStop}%
\bibitem [{\citenamefont {{Vallisneri}}\ \emph {et~al.}(2008)\citenamefont
  {{Vallisneri}}, \citenamefont {{Crowder}},\ and\ \citenamefont
  {{Tinto}}}]{2008CQGra..25f5005V}%
  \BibitemOpen
  \bibfield  {author} {\bibinfo {author} {\bibfnamefont {M.}~\bibnamefont
  {{Vallisneri}}}, \bibinfo {author} {\bibfnamefont {J.}~\bibnamefont
  {{Crowder}}}, \ and\ \bibinfo {author} {\bibfnamefont {M.}~\bibnamefont
  {{Tinto}}},\ }\href {\doibase 10.1088/0264-9381/25/6/065005} {\bibfield
  {journal} {\bibinfo  {journal} {Class. Quant. Grav.}\ }\textbf {\bibinfo
  {volume} {25}},\ \bibinfo {eid} {065005} (\bibinfo {year}
  {2008})}\BibitemShut {NoStop}%
\bibitem [{\citenamefont {Williams}\ and\ \citenamefont
  {Rasmussen}(2006)}]{gpbook2006}%
  \BibitemOpen
  \bibfield  {author} {\bibinfo {author} {\bibfnamefont {C.~K.~I.}\
  \bibnamefont {Williams}}\ and\ \bibinfo {author} {\bibfnamefont {C.~E.}\
  \bibnamefont {Rasmussen}},\ }\href@noop {} {\emph {\bibinfo {title} {Gaussian
  processes for machine learning}}}\ (\bibinfo  {publisher} {MIT press},\
  \bibinfo {address} {Cambridge, MA},\ \bibinfo {year} {2006})\BibitemShut
  {NoStop}%
\bibitem [{\citenamefont {{van Haasteren}}\ and\ \citenamefont
  {{Levin}}(2013)}]{2013MNRAS.428.1147V}%
  \BibitemOpen
  \bibfield  {author} {\bibinfo {author} {\bibfnamefont {R.}~\bibnamefont {{van
  Haasteren}}}\ and\ \bibinfo {author} {\bibfnamefont {Y.}~\bibnamefont
  {{Levin}}},\ }\href {\doibase 10.1093/mnras/sts097} {\bibfield  {journal}
  {\bibinfo  {journal} {MNRAS}\ }\textbf {\bibinfo {volume} {428}},\ \bibinfo
  {pages} {1147} (\bibinfo {year} {2013})}\BibitemShut {NoStop}%
\bibitem [{\citenamefont {{van Haasteren}}\ and\ \citenamefont
  {{Vallisneri}}(2014)}]{2014PhRvD..90j4012V}%
  \BibitemOpen
  \bibfield  {author} {\bibinfo {author} {\bibfnamefont {R.}~\bibnamefont {{van
  Haasteren}}}\ and\ \bibinfo {author} {\bibfnamefont {M.}~\bibnamefont
  {{Vallisneri}}},\ }\href {\doibase 10.1103/PhysRevD.90.104012} {\bibfield
  {journal} {\bibinfo  {journal} {\prd}\ }\textbf {\bibinfo {volume} {90}},\
  \bibinfo {eid} {104012} (\bibinfo {year} {2014})}\BibitemShut {NoStop}%
\bibitem [{\citenamefont {Topçu}(1979)}]{topcu1979}%
  \BibitemOpen
  \bibfield  {author} {\bibinfo {author} {\bibfnamefont {A.}~\bibnamefont
  {Topçu}},\ }\href@noop {} {\emph {\bibinfo {title} {A contribution to the
  systematic analysis of finite element structures using the force method}}}\
  (\bibinfo  {publisher} {Essen University},\ \bibinfo {year} {1979})\ \bibinfo
  {note} {doctoral dissertation}\BibitemShut {NoStop}%
\bibitem [{\citenamefont {Kaneko}\ \emph {et~al.}(1982)\citenamefont {Kaneko},
  \citenamefont {Lawo},\ and\ \citenamefont {Thierauf}}]{kaneko1982}%
  \BibitemOpen
  \bibfield  {author} {\bibinfo {author} {\bibfnamefont {I.}~\bibnamefont
  {Kaneko}}, \bibinfo {author} {\bibfnamefont {M.}~\bibnamefont {Lawo}}, \ and\
  \bibinfo {author} {\bibfnamefont {G.}~\bibnamefont {Thierauf}},\ }\href@noop
  {} {\bibfield  {journal} {\bibinfo  {journal} {Int. J. Num. Methods Eng.}\
  }\textbf {\bibinfo {volume} {18}},\ \bibinfo {pages} {1469} (\bibinfo {year}
  {1982})}\BibitemShut {NoStop}%
\bibitem [{\citenamefont {Berry}\ \emph {et~al.}(1985)\citenamefont {Berry},
  \citenamefont {Heath}, \citenamefont {Kaneko}, \citenamefont {Lawo},
  \citenamefont {Plemmons},\ and\ \citenamefont {Ward}}]{berry1985}%
  \BibitemOpen
  \bibfield  {author} {\bibinfo {author} {\bibfnamefont {M.}~\bibnamefont
  {Berry}}, \bibinfo {author} {\bibfnamefont {M.}~\bibnamefont {Heath}},
  \bibinfo {author} {\bibfnamefont {I.}~\bibnamefont {Kaneko}}, \bibinfo
  {author} {\bibfnamefont {M.}~\bibnamefont {Lawo}}, \bibinfo {author}
  {\bibfnamefont {R.}~\bibnamefont {Plemmons}}, \ and\ \bibinfo {author}
  {\bibfnamefont {R.}~\bibnamefont {Ward}},\ }\href@noop {} {\bibfield
  {journal} {\bibinfo  {journal} {Numerische Mathematik}\ }\textbf {\bibinfo
  {volume} {47}},\ \bibinfo {pages} {483} (\bibinfo {year} {1985})}\BibitemShut
  {NoStop}%
\bibitem [{\citenamefont {Anderson}\ \emph {et~al.}(1999)\citenamefont
  {Anderson} \emph {et~al.}}]{laug}%
  \BibitemOpen
  \bibfield  {author} {\bibinfo {author} {\bibfnamefont {E.}~\bibnamefont
  {Anderson}} \emph {et~al.},\ }\href@noop {} {\emph {\bibinfo {title}
  {{LAPACK} Users' Guide}}},\ \bibinfo {edition} {3rd}\ ed.\ (\bibinfo
  {publisher} {SIAM},\ \bibinfo {address} {Philadelphia, PA},\ \bibinfo {year}
  {1999})\BibitemShut {NoStop}%
\bibitem [{\citenamefont {Golub}\ and\ \citenamefont
  {Van~Loan}(1996)}]{golub1996}%
  \BibitemOpen
  \bibfield  {author} {\bibinfo {author} {\bibfnamefont {G.~H.}\ \bibnamefont
  {Golub}}\ and\ \bibinfo {author} {\bibfnamefont {C.~F.}\ \bibnamefont
  {Van~Loan}},\ }\href@noop {} {\emph {\bibinfo {title} {Matrix
  Computations}}}\ (\bibinfo  {publisher} {Johns Hopkins University Press},\
  \bibinfo {address} {Baltimore and London},\ \bibinfo {year}
  {1996})\BibitemShut {NoStop}%
\bibitem [{\citenamefont {Dang}\ \emph {et~al.}(2017)\citenamefont {Dang},
  \citenamefont {Ling},\ and\ \citenamefont {Maciejowski}}]{dang2017}%
  \BibitemOpen
  \bibfield  {author} {\bibinfo {author} {\bibfnamefont {T.~V.}\ \bibnamefont
  {Dang}}, \bibinfo {author} {\bibfnamefont {K.~V.}\ \bibnamefont {Ling}}, \
  and\ \bibinfo {author} {\bibfnamefont {J.}~\bibnamefont {Maciejowski}},\
  }\href@noop {} {\bibfield  {journal} {\bibinfo  {journal}
  {IFAC-PapersOnLine}\ }\textbf {\bibinfo {volume} {50}},\ \bibinfo {pages}
  {13170} (\bibinfo {year} {2017})}\BibitemShut {NoStop}%
\bibitem [{\citenamefont {Eberly}(2004)}]{eberly2004}%
  \BibitemOpen
  \bibfield  {author} {\bibinfo {author} {\bibfnamefont {W.}~\bibnamefont
  {Eberly}},\ }in\ \href {\doibase 10.1145/1005285.1005305} {\emph {\bibinfo
  {booktitle} {Proceedings of the 2004 International Symposium on Symbolic and
  Algebraic Computation}}}\ (\bibinfo  {publisher} {ACM},\ \bibinfo {address}
  {New York},\ \bibinfo {year} {2004})\ p.\ \bibinfo {pages}
  {127–134}\BibitemShut {NoStop}%
\bibitem [{\citenamefont {Coakley}\ \emph {et~al.}(2011)\citenamefont
  {Coakley}, \citenamefont {Rokhlin},\ and\ \citenamefont
  {Tygert}}]{coakley2011}%
  \BibitemOpen
  \bibfield  {author} {\bibinfo {author} {\bibfnamefont {E.}~\bibnamefont
  {Coakley}}, \bibinfo {author} {\bibfnamefont {V.}~\bibnamefont {Rokhlin}}, \
  and\ \bibinfo {author} {\bibfnamefont {M.}~\bibnamefont {Tygert}},\ }\href
  {\doibase 10.1137/090779656} {\bibfield  {journal} {\bibinfo  {journal} {SIAM
  J. Sci. Comp.}\ }\textbf {\bibinfo {volume} {33}},\ \bibinfo {pages} {849}
  (\bibinfo {year} {2011})}\BibitemShut {NoStop}%
\bibitem [{\citenamefont {Kressner}\ and\ \citenamefont
  {Šušnjara}(2017)}]{kressner2017}%
  \BibitemOpen
  \bibfield  {author} {\bibinfo {author} {\bibfnamefont {D.}~\bibnamefont
  {Kressner}}\ and\ \bibinfo {author} {\bibfnamefont {A.}~\bibnamefont
  {Šušnjara}},\ }\href {\doibase 10.1137/16M1087278} {\bibfield  {journal}
  {\bibinfo  {journal} {SIAM J. Matrix Analysis Appl.}\ }\textbf {\bibinfo
  {volume} {38}},\ \bibinfo {pages} {984} (\bibinfo {year} {2017})}\BibitemShut
  {NoStop}%
\end{thebibliography}%

\end{document}